\documentclass[twocolumn,showpacs,preprintnumbers,amsmath,amssymb]{revtex4}
\usepackage{graphicx}% Include figure files
\usepackage{dcolumn}% Align table columns on decimal point
\usepackage{bm}% bold math

\begin{document}

%\preprint{APS/123-QED}

\title{Novel quantum phases of dipolar Bose gases in optical lattices}
\author{S. Yi$^1$, T. Li$^2$, and C. P. Sun$^1$}

\affiliation{$^1$Institute of Theoretical Physics, Chinese Academy
of Sciences, Beijing, 100080, P. R. China}

\affiliation{$^2$Department of Physics, Renmin University of China,
Beijing, 100872, P.R.China}

\begin{abstract}
We investigate the quantum phases of polarized dipolar Bosons loaded into a two-dimensional square and three-dimensional cubic optical lattices. We show that the long-range and anisotropic nature of the dipole-dipole interaction induces a rich variety of quantum phases, including the supersolid and striped supersolid phases in 2D lattices, and the layered supersolid phase in 3D lattices.
\end{abstract}

\date{\today}

\maketitle

The theoretical prediction \cite{jaksch} and experimental realization~\cite{bloch} of the superfluid (SF) to Mott insulating (MI) transition in atomic system have triggered tremendous research activities on ultracold atoms in a optical lattice. The unprecedented experimental control over both atomic and optical lattice parameters makes it an ideal platform for studying the physics of strongly correlated quantum system. The physics of atomic gases in an optical lattice is captured by the on-site Bose-Hubbard model~\cite{jaksch}, however, it has the limitation of lacking off-site density-density correlation, which is essential for the emergence of density wave (DW) order and more intriguingly the supersolid (SS) phase~\cite{ss1,ss2,goral,ss3,scarola}. Recently, it was proposed that the off-site density-density couplings can be realized by the bosons in higher bands of optical lattices~\cite{scarola}, by interaction mediated through fermions in a mixture of bosonic and fermionic atoms~\cite{buchler}, and more directly, by the long-range dipole-dipole interaction between dipolar bosons~\cite{goral}. Among these, dipolar interaction has the advantage of high tunability and experimental accessibility, especially with the recently achieved chromium condensate~\cite{pfau} and ultracold polar molecules in a 3D optical lattice~\cite{ospel}.

So far, ultracold dipolar Bose gases have attracted significant theoretical interest, for the dipole-dipole interaction gives rise to new phenomena in Bose-Einstein condensates~\cite{dipbec,damski,xie,spintxt} and provides new schemes for quantum computing~\cite{qucomp}. In particular, when loaded into optical lattice potential, the dipolar interaction may induce exotic magnetic orderd if the dipole moments are free to rotate in space~\cite{polarm,qumag,mich}.

In the present work, we map out the phase diagrams of polarized dipolar bosons in 2D square and 3D cubic optical lattices. For a 2D lattice on $xy$ plane, the dipole moments may point either to $z$- or $y$-axis, the former corresponds a {\em 2D isotropic model} as the dipolar interaction is isotropic on the lattice plane, while the latter is referred to as the {\em 2D anisotropic model} for the anisotropic dipolar interaction on $xy$ plane. For 3D lattices, we always assume that the dipole moments point to $z$ direction. As we shall show below, the long-range and anisotropic nature of the dipolar interaction give rise to extremely rich quantum phases in the lattice systems. In addition to the SS and checkerboard phases found in Ref.~\cite{goral}, the striped supersolid (SSS) phases are also observed in 2D anisotropic model. More remarkably, by tuning the sign of the dipole-dipole interaction to negative, we discover the layered supersolid (LSS) phase in 3D lattices.

In the presence of dipole-dipole interaction, the Hamiltonian for the extended Bose-Hubbard model reads~\cite{goral}
\begin{eqnarray}
H\!\!&=&\!\!-t\sum_{\langle i,j\rangle} (b_i^\dag  b_j+b_j^\dag  b_i)-\mu\sum_i
n_i+\frac{U_0}{2}\sum_i n_i( n_i-1)\nonumber\\
&&+\frac{1}{2}\sum_i U_{\rm dd}^{ii}n_i(
n_i-1)+\frac{1}{2}\sum_{i\neq j} U_{\rm dd}^{ij}n_i n_j,\label{ham}
\end{eqnarray}
where $b_i^\dag$ is the boson creation operator at site $i$ and $n_i=b_i^\dag b_i$ is the corresponding particle number operator. $t$ is the hopping matrix element between nearest neighbors, $U_0>0$ is the on-site Hubbard repulsion due to $s$-wave scattering, and throughout this paper, the chemical potential $\mu$ is fixed at $0.4U_0$.

The last line of Eq. (\ref{ham}) describes the dipolar interaction in the lattice system with coupling parameters being expressed as
\begin{eqnarray}
U_{\rm dd}^{ij}=f c_{\rm dd} {\cal D}_{ij},
\end{eqnarray}
where $f$ is a factor continuously tunable between $-\frac{1}{2}$ and $1$ via a fast rotating orienting field~\cite{giov}. As shown below, nontrivial quantum phases appear for negative $f$ values. $c_{\rm dd}=d^2/(4\pi\varepsilon_0)$ or $\mu_0d^2/(4\pi)$ for, respectively, the electric or magnetic dipoles, with $d$ being the dipole moment and $\varepsilon_0$ ($\mu_0$) being the vacuum permittivity (permeability). \begin{eqnarray}
{\cal D}_{ij}=\int d{\mathbf r}d{\mathbf r}' |w({\mathbf r}-{\mathbf r}_i)|^2\frac{1-3\cos^2\theta}{|{\mathbf r}-{\mathbf r}'|^3}|w({\mathbf r}'-{\mathbf r}_j)|^2,
\end{eqnarray}
where $w({\mathbf r}-{\mathbf r}_i)$ is the localized (on site $i$) Wannier function of the lowest energy band and $\theta$ is the angle formed by the dipole moment and the vector $(\mathbf r-{\mathbf r}')$. Since ${\cal D}_{ij}$ only depends on ${\mathbf r}_i-{\mathbf r}_j=a(l_x\hat{\mathbf x}+l_y\hat{\mathbf y}+l_z\hat{\mathbf z})$ with $a$ being lattice constant and $l_{x,y,z}$ being integers, it can be equivalently denoted as ${\cal D}_{(l_xl_yl_z)}$.

In principal, $t$, $U_0$, and $\{U_{\rm dd}^{ij}\}$ are all determined by the optical lattice parameters~\cite{goral}. However, to obtain a complete phase diagram, we shall allow them to change independently except that the set of $\{U_{\rm dd}^{ij}\}$ have to be calculated consistently for a given lattice geometry. For simplicity, we calculate ${\cal D}_{(l_xl_yl_z)}$ using a spherical Gaussian function of width $\sigma$, $w({\mathbf r}-{\mathbf r}_i)=\pi^{-3/4}\sigma^{-3/2}e^{-({\mathbf r}-{\mathbf r}_i)^2/(2\sigma^2)}$, for which ${\cal D}_{(000)}$ vanish~\cite{gauss}. All results presented in this work are obtained with ${\cal D}_{(l_xl_yl_z)}$ corresponding to $(\sigma,a)=(1,3)$. Similar numerical results are found with other $(\sigma,a)$ combinations. For practical purpose, we choose to truncate $l_{x,y,z}$ according to $-l_{\rm max}\leq l_{x,y,z}\leq l_{\rm max}$ with $l_{\rm max}=2$. We have also performed simulations using $l_{\rm max}$ up to $6$ and $3$ for, respectively, 2D and 3D lattices, which yield similar qualitative pictures. To simplify the notation, we define a dimensionless parameter
\begin{eqnarray}
\gamma\equiv f c_{\rm dd}(4\pi\hbar^2a_{\rm sc}/M)^{-1}\nonumber
\end{eqnarray}
to measure the relative strength of the dipolar coupling, where $a_{\rm sc}$ is the $s$-wave scattering length and $M$ is the mass of the atom. It can be easily estimated that $\gamma\approx 0.033f$ for $^{52}$Cr atom and $\gamma\approx5.68f$ for a typical polar molecule~\cite{molecule}. Finally, the mean-field ground state of Hamiltonian (\ref{ham}) is obtained using Gutzwiller ansatz~\cite{jaksch,goral} on lattices up to $96\times 96$ and $24\times 24\times 24$ sites with periodic boundary conditions. Similar results are expected for even large number of lattice sites. The Gutzwiller ansatz has been widely used to study the ground state and the dynamics of cold bosons in optical lattices\cite{jaksch,goral,damski}, as it can capture qualitatively the quantum phases, especially in 3D lattices~\cite{zwerger}.

\begin{figure}
\centering
\includegraphics[width=2.5in]{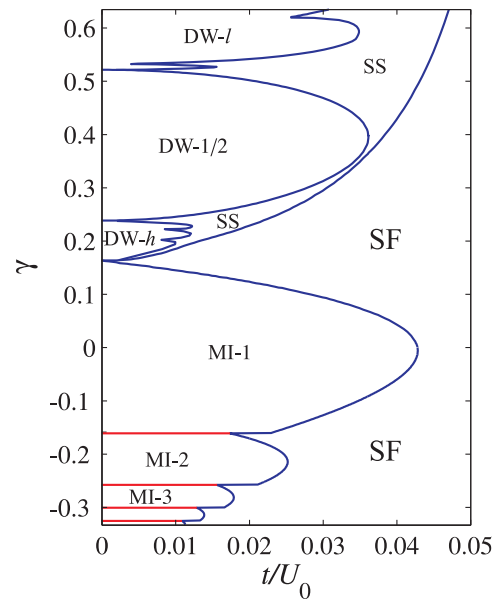}
\caption{(Color online) Phase diagram in $t$-$\gamma$ parameter plane for 2D isotropic model obtained on a $12\times12$ lattice. Various MI-$\rho$ phases are distinguished by the occupation number $\rho$. DW-$1/2$ denotes the checkerboard phase. In DW-$h$ (DW-$l$) region, there exist many small DW lobes with the mean densities higher (lower) than $1/2$. The phase transitions between various insulating phases are of first order, while all other transitions are of second order. This observation also applies to Figs. \ref{dip2daniphase} and \ref{dip3dphase} (a).}
\label{dip2disophase}
\end{figure}

{\em 2D isotropic model}. -- As shown in Fig. \ref{dip2disophase}, only MI and SF phases appear when $\gamma$ is negative, which is similar to the on-site Bose-Hubbard model with $|\gamma|$ playing effectively the role of the chemical potential. The boundary between two adjacent MI phases is a horizontal line, whose position can be determined analytically as follows: the energy per site for an MI phase is
\begin{eqnarray}
{\cal E}/U_0=\rho(\rho-1)/2-\mu\rho/U_0+\gamma\varepsilon_{\rm dd}\rho^2,\label{engmott}
\end{eqnarray}
where $\rho$ is the mean particle density and $\gamma\varepsilon_{\rm dd}$ is the dipolar interaction energy per site for MI-$1$ phase. The boundary between MI-$\rho$ and MI-$(\rho+1)$ phases then locates at
$\gamma=(\mu/U_0-\rho)/[\varepsilon_{\rm dd}(2\rho+1)]$. Furthermore, a MI phase is stable only if
\begin{eqnarray}
\gamma\varepsilon_{\rm dd}+1/2>0,\label{stability}
\end{eqnarray}
hence the system collapses for $\gamma\lesssim -0.41$, in very good agreement with our numerical result. This stability criterion also applies roughly to the SF phase.
%We point out that, unlike the collapse mechanism arising from on-site attraction~\cite{goral}, here it is induced by the long-ranged attractive interaction.

For positive $\gamma$, the SS and DW phases emerge in addition to  the usual SF and MI phases. In particular, the large lobe denoted as DW-$1/2$ corresponds to the checkerboard phase, where sites are alternating empty and singly occupied. We note that the checkerboard and SF phases are not connected directly, in accord with the Landau theory of phase transition, namely, phases with distinct symmetry breaking patterns can not be connected with each other continuously. Both DW-$h$ and -$l$ regions contain many small DW lobes with different mean densities, as these DW lobes are very sensitive to $t$, $\gamma$, and the size of lattice, it is very difficult to map them out in detail.

From above discussions, it is clear that the repulsive dipolar interaction ($\gamma>0$) with sufficient strength breaks the translational symmetry, while the attractive one ($\gamma<0$) always preserves it. The symmetry breaking and preserving property of the dipolar interactions can be understood intuitively as follows: supposing that site $i$ is occupied, it is energetically favorable that its neighboring sites are equally (less) populated if the dipole-dipole interaction is attractive (repulsive).

\begin{figure}
\centering
\includegraphics[width=2.5in]{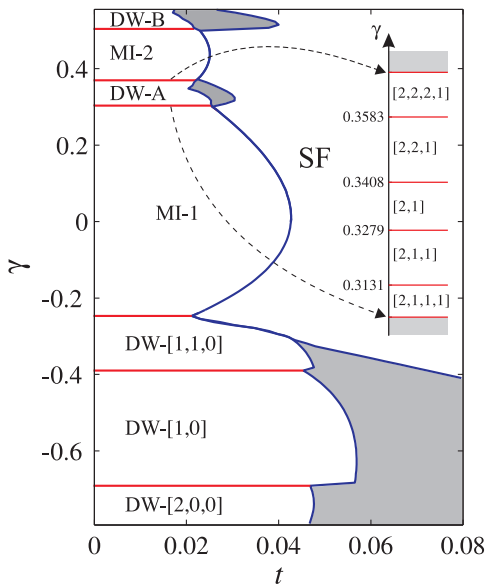}
\caption{(Color online) Same as Fig. \ref{dip2disophase} except for 2D anisotropic model. The DW phases for $\gamma<0$ are distinguished by their density distributions. For $\gamma>0$, both DW-A and -B regions contain fine structures. The shaded regions represent the SSS phases. The inset shows the zoom-in plot of DW-A region, the numbers on the left hand side of the $\gamma$ axis mark the boundaries between two DW phases, and the density distributions are specified on other side. DW-B has a similar structure except that the occupation number on each site is increased by 1.} \label{dip2daniphase}
\end{figure}

{\em 2D anisotropic model}. -- As dipole moments are polarized along $y$-axis, the dipolar interaction along $y$ ($x$) direction is attractive (repulsive) for positive $\gamma$, while for negative $\gamma$, the opposite is true. Based on the relation between dipolar interaction and translational symmetry, we expect to find stripe ordering here. The phase diagram presented in Fig. \ref{dip2daniphase} indeed confirms our conjecture. Considering, for example, the insulating phases in Fig. \ref{dip2daniphase}, the particle density along the repulsive direction becomes periodically modulated for sufficiently large $|\gamma|$, while it remains to be a constant along the attractive direction. The structure of a striped insulating phase is then completely specified by the particle densities over a single modulation period as $[\rho_1,\rho_2,\ldots,\rho_m]$.

There exists a subtle difference between the dipolar interactions corresponding to negative and positive $\gamma$: for a uniform density distribution, the overall dipolar interaction is attractive (repulsive) for the positive (negative) $\gamma$. Consequently, as the dipolar coupling grows, the mean density generally increases for positive $\gamma$, while decreases for negative $\gamma$. The DW-[2,0,0] phase, on the other hand, is an exception, as it has higher mean density compared to DW-[1,0] phase. This can be easily understood from its density distribution, for which the attractive dipolar interaction within each lattice row is maximized, while the repulsive one between rows is minimized. As we further increase $|\gamma|$, the system collapses for roughly $\gamma\gtrsim 0.63$ and $\gamma\lesssim-1.27$. More interestingly, the shaded regions, to the right of the striped solid phases, represent the SSS phase with a density distribution resembling their insulting counterparts. Recently, the stripe ordered SS phase was also predicted in the $p$-band Bose-Hubbard model of 2D triangular lattices~\cite{wuc}.

\begin{figure}
\centering
\includegraphics[width=2.4in]{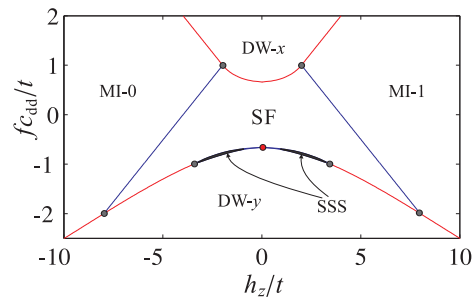}
\caption{(Color online) Mean-field phase diagram for spin-$\frac{1}{2}$ model Eq. (\ref{hxxz}) with nearest neighbor coupling. In MI-0 phase, lattice sites are not occupied. DW-$x$ (DW-$y$) denotes the striped solid phase with density modulation along $x$ ($y$) direction. The shaded regions represent the SSS phases.} \label{anixxz}
\end{figure}

To gain more insight into the stripe ordering, we re-exam the Hamilton (\ref{ham}) in the hard-core limit, where it maps onto a spin-$\frac{1}{2}$ $X\!X\!Z$ model
\begin{eqnarray}
H=-t\sum_{\langle i,j\rangle}(s_i^+s_j^-+h.c.)+\frac{1}{2}\sum_{i\neq j}U_{dd}^{ij}s_i^zs_j^z-h_z\sum_is_i^z\label{hxxz}
\end{eqnarray}
with $h_z=\mu-\gamma\epsilon_{\rm dd}/2$. Here the ferromagnetic ordering in $xy$ plane corresponds to the superfluidity in Eq. (\ref{ham}), and the modulation in $s_i^z$ corresponds to the ordering of the density. For simplicity, we only take into account the dipolar interaction between nearest neighboring (nn) sites. Figure \ref{anixxz} shows the phase diagram obtained through numerical optimization. Compared to Fig. \ref{dip2daniphase}, the SSS regions here are rather small. Since inclusion of the next-nearest-neighboring (nnn) coupling does not help much, we conclude that the soft-core nature of Eq. (\ref{ham}) should be responsible for the large stable SSS phases in Fig. \ref{dip2daniphase}. This also explains the absence of SSS phase for $f>0$.
For isotropic model, nnn coupling is the key factor for the presence of SSS phase. It not only provides frustration necessary for SS order, but is also responsible for the stripe ordering at large nnn coupling strength~\cite{xxz}. Here, however, due to the anisotropic nature of the long-range interaction, the nn coupling alone is sufficient for generating the SSS phase.

\begin{figure}
\centering
\includegraphics[width=3.in]{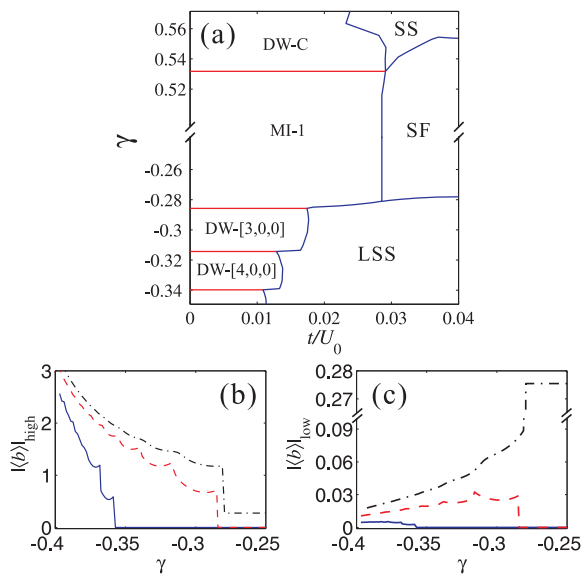}
\caption{(Color online) (a) Mean-field phase diagram for dipolar bosons in 3D cubic lattice obtained on a $6\times6\times6$ lattice. The lower panels show the $\gamma$ dependence of superfluid parameters on high (b) and low (c) density layers for $t/U_0=0.01$ (solid lines), $0.02$ (dashed lines), and $0.03$ (dash-dotted lines).} \label{dip3dphase}
\end{figure}

{\em 3D lattices}. -- In Fig. \ref{dip3dphase} (a), we plot the phase diagram for dipolar bosons in a 3D lattice. A special feature we note immediately is that the right boundary of MI-1 phase is a vertical line. This is because our cut-off scheme on $l_{x,y,z}$ makes the dipolar interaction energy vanishes for a uniform density distribution on a 3D cubic lattice. For positive $\gamma$, the dipolar interaction is repulsive on the lattice planes perpendicular to $z$-axis, therefore DW order emerges on each lattice layer for sufficiently large $\gamma$, in analogy to the 2D isotropic model. However, as the dipolar interaction is attractive along $z$-axis, the density distributions on different lattice layers are identical. The detailed structure of DW-C region is rather complicated. In general, the mean density first decreases with $\gamma$, then increases slightly until the system collapses at $\gamma\approx0.63$. The SS phase has a similar density structure to that of DW-C phase.

For negative $\gamma$, one can naturally expect the layered phases due to the attractive dipolar interaction on lattice layer. In other words, for a layered phase, the particle density within each layer is a constant while periodically modulated along $z$-axis. In Fig. \ref{dip3dphase} (a), the layered insulating phase are specified by the densities on lattice layers over a single modulation period. Unlike the 2D case, the only density modulation period we found is 3 even for a lattice whose size is not a multiple of 3. Same as the 2D isotropic case, the onset of instability occurs at $\gamma\approx-0.41$.

The structure of the LSS phase is similar to that of a layered solid phase: each modulation period contains a layer with higher density and two layers with equal lower densities. Figure \ref{dip3dphase} (b) and (c) show the $\gamma$ dependence of the SF order parameter $|\langle b\rangle|$ on, respectively, the high and low density layers. In general, SF parameter increases with $|\gamma|$ on high density layers, while decreases on low density layers. In the strong dipolar interaction limit, $|\langle b\rangle|_{\rm high}$ can be two orders of magnitude larger than $|\langle b\rangle|_{\rm low}$. If we treat the low density layers as charge reservoir unit, then the resulting structure is quite similar to that of a high-$T_c$ cuprate superconductor.

In conclusion, we have studied the quantum phases of dipolar bosons in optical lattice. We shown that, due to the long-range and anisotropic nature of dipole-dipole interaction and its tunability, the achievable phases in this system are extremely rich, which makes dipolar bosons in a lattice an ideal candidate for studying the exotic phases in strongly correlated quantum system. The experimental detection of SSS and LSS phases in chromium atoms relies on tuning the scattering length to below 10 Bohr radius, while they should be observable in typical polar molecule systems.

This work is supported by the NSFC and the National 973 project of China. S.Y. is also supported by the ``BaiRen" program of CAS. We thank Professor H. Pu for a critical reading of the manuscript. The helpful discussions with Professor Jinwu Ye and the correspondence with Professor L. Santos on numerical calculation are acknowledged.


\begin{thebibliography}{}
\bibitem{gauss} Even for a 2D lattice, the spherically symmetrical $w({\mathbf r}-{\mathbf r}')$ can still be realized by applying a harmonic trap with appropriate trapping frequency along $z$ direction.
\bibitem{molecule} For chromium atoms, $d=6\mu_B$ and $a_{\rm sc}=112a_{B}$; For typical polar molecules, $d=0.5$ Debye, $a_{\rm sc}=100a_{B}$, and $M$ is around 100 atomic number.
\end{thebibliography}
\end{document}